\documentclass{emulateapj}
\newcommand{\ofo}{0509$-$67.5}
\newcommand{\ofn}{0519$-$69.0}
 
\usepackage{color}

\slugcomment{Draft Version \today}

\shorttitle{PROGENITORS OF LMC SNe}
\shortauthors{Badenes et al.}

\begin{document}

\title{THE STELLAR ANCESTRY OF SUPERNOVAE IN THE MAGELLANIC CLOUDS - I. THE MOST RECENT SUPERNOVAE IN THE LARGE MAGELLANIC CLOUD}

\author{Carles Badenes\altaffilmark{1,2}, Jason Harris\altaffilmark{3}, Dennis Zaritsky\altaffilmark{4} and Jos\'e Luis Prieto\altaffilmark{5}}

\altaffiltext{1}{Department of Astrophysical Sciences, Princeton University, Peyton Hall, Ivy Lane, Princeton, NJ 08544;
  badenes@astro.princeton.edu}

\altaffiltext{2}{\textit{Chandra} Fellow}

\altaffiltext{3}{National Optical Astronomy Observatory, 950 North Cherry Ave., Tucson, AZ 85719; jharris@noao.edu}

\altaffiltext{4}{Steward Observatory, 933 North Cherry Ave., Tucson, AZ 85721; dzaritsky@as.arizona.edu}

\altaffiltext{5}{Department of Astronomy, Ohio State University, McPherson Laboratory, 140 W. 18th Avenue. Columbus, OH
  43210; prieto@astronomy.ohio-state.edu}

\begin{abstract}
  We use the star formation history map of the Large Magellanic Cloud recently published by Harris \& Zaritsky to study
  the sites of the eight smallest (and presumably youngest) supernova remnants in the Cloud: SN 1987A, N158A, N49, and
  N63A (core collapse remnants), \ofo, \ofn, N103B, and DEM L71 (Type Ia remnants). The local star formation histories
  provide unique insights into the nature of the supernova progenitors, which we compare with the properties of the
  supernova explosions derived from the remnants themselves and from supernova light echoes. We find that all the core
  collapse supernovae that we have studied are associated with vigorous star formation in the recent past. In the case
  of SN 1987A, the time of the last peak of star formation (12 Myr) matches the lifetime of a star with the known mass
  of its blue supergiant progenitor ($\sim20\,\mathrm{M_{\odot}}$). More recent peaks of star formation can lead to
  supernovae with more massive progenitors, which opens the possibility of a Type Ib/c origin for SNRs N158A and
  N63A. Stars more massive than $21.5\, \mathrm{M_{\odot}}$ are very scarce around SNR N49, implying that the magnetar
  SGR 0526$-$66 in this SNR was either formed elsewhere or came from a progenitor with a mass well below the
  $30\,\mathrm{M_{\odot}}$ threshold suggested in the literature. Three of our four Ia SNRs are associated with old,
  metal poor stellar populations. This includes SNR \ofo, which is known to have been originated by an extremely bright
  Type Ia event, and yet is located very far away from any sites of recent star formation, in a population with a mean
  age of $7.9$ Gyr. The Type Ia SNR N103B, on the other hand, is associated with recent star formation, and might have
  had a relatively younger and more massive progenitor with substantial mass loss before the explosion. We discuss these
  results in the context of our present understanding of core collapse and Type Ia supernova progenitors.
\end{abstract}


\section{INTRODUCTION}
\label{sec:Intro}

The identification of the progenitor stars of supernova (SN) explosions is one of the central problems of stellar
astrophysics. In the case of core collapse supernovae (CC SNe: Types II, Ib, Ic, and derived subtypes) the progenitors
are known to be massive ($M>8\,\mathrm{M_{\odot}}$) stars whose inner cores collapse to a neutron star or a black
hole. In a few cases, it has been possible to constrain the properties of the progenitor star using pre-explosion images
or the turn-off masses of compact clusters
\citep{crockett08:SN2007gr_progenitor,smartt08:death_massive_stars_I,vinko09:SN2004dj_progenitor} , but there are still
many open issues regarding which stars lead to specific subtypes of CC SNe \citep[for an extended discussion and a
complete set of references, see][]{smartt08:death_massive_stars_I,kochanek08:survey_nothing}. In the case of
thermonuclear (Type Ia) SNe, the situation is much more complex. Although a CO white dwarf (WD) in some kind of binary
is almost certainly the exploding star, the exact nature of the progenitor system has never been firmly established,
either theoretically or observationally \citep[see][and references
therein]{maoz08:fraction_intermediate_stars_Ia_progenitors}.

When direct identifications are not possible, the properties of the progenitors can be constrained using the stellar
populations around the exploding stars. A number of studies have done this for SNe in nearby galaxies
\citep[e.g.][]{hamuy00:Environment_effects_SNIa,sullivan06:SNIa_rate_host_SFR,
  aubourg07:massive_stars_SNIa,modjaz08:GRB_metallicities,prieto08:SN_Progenitors_Metallicities,
  gallagher08:SNIa_Early_Type_Galaxies}, but this approach has important limitations. First, the available information,
be it photometric \citep[e.g.][]{sullivan06:SNIa_rate_host_SFR} or spectral
\citep[e.g.][]{gallagher08:SNIa_Early_Type_Galaxies}, is usually integrated over the entire host galaxy, although local
measurements at the SN sites have been made for a small number of objects
\citep[e.g.][]{modjaz08:GRB_metallicities}. This effectively ignores the metallicity and stellar age gradients that must
be present in the host. Second, even in surveys that work with complete host spectra, the stellar populations are not
resolved. Among other things, this means that the stellar light is weighted by luminosity, which can conceal many
important properties of the stellar populations. In practice, the information that can be obtained from this kind of
observations is restricted to average metallicities and ages, unless sophisticated fitting techniques are used to
extract the star formation history (SFH) of the host \citep[see][]{aubourg07:massive_stars_SNIa}. Ideally, one would
want to study \textit{resolved} stellar populations associated with SN progenitors. The information that can be obtained
in this way is much more detailed and reliable, but it requires focusing on very nearby SNe.

The present work is the first in a series of papers aimed at constraining the fundamental properties of CC and Ia SN
progenitors in the Magellanic Clouds by examining the stellar populations at the locations of the supernova remnants
(SNRs) left behind by the explosions. To do this, we take advantage of the large amount of observational data
accumulated on the stellar populations of the Clouds, in particular the star formation history (SFH) maps published by
\citet{harris04:SMC_SFH} for the SMC and \citet{harris08:LMC_SFH} (henceforth, HZ09) for the LMC. To identify the sites
of recent SNe, we rely on the extensively observed population of SNRs in the MCs \citep{williams99:LMC_SNR_Atlas}. Much
information about the SN explosions can be extracted from the observations of SNRs of both Ia and CC origin
\citep{badenes03:xray,chevalier05:young_cc_SNRs}, and in some cases this information can be complemented by light echoes
from the SNe themselves \citep{rest05:LMC_light_echoes,rest08:0509}. In this first installment, we focus on the eight
youngest SNRs in the LMC: SN1987A, N158A, N63A, and N49 (CC SNRs); \ofo, \ofn, N103B and DEM L71 (Ia SNRs).

This paper is organized as follows. In \S~\ref{sec:Targets} we describe the criteria that have led to the selection of
our eight target SNRs. In \S~\ref{sec:SNtoSNR} we review their types and the characteristics of the parent SNe that can
be inferred from their observational properties. In \S~\ref{sec:Map} we review the fundamental details of the SFH map of
the LMC presented in HZ09. In \S~\ref{sec:Relevance} we discuss the relevance that the local SFH has for the properties
of the SN progenitors, given our knowledge about the global SFH and the stellar dynamics of the LMC. In \S~\ref{sec:SFH}
we examine the local SFHs for the target SNRs, with specific comments relating each SFH to the SNe that originated the
SNRs. In \S~\ref{sec:Disc} we discuss the impact that our findings have in the context of our current understanding of
CC and Ia SN progenitors. Finally, in \S~\ref{sec:Conc} we present our conclusions and we outline some avenues for
future research.

\section{TARGET SELECTION}
\label{sec:Targets}

We will focus on young SNRs because they are usually ejecta-dominated and still contain a great deal of information
about their parent SNe - in particular, the risk of mistyping young CC and Ia SNRs is minimal (see
\S~\ref{sec:SNtoSNR}). Identifying the youngest SNRs in a given set, however, is not trivial. Among the SNRs in the LMC,
only one has a known age (SN 1987A), and only three (\ofo, \ofn, and N103B) have more or less accurate age estimates
from light echoes \citep{rest05:LMC_light_echoes}. In the absence of consistent age estimates for all objects, size is
the best criterion to select the youngest ones. Much information about the SNR population in the LMC can be found in the
\textit{ROSAT} atlas by \citet{williams99:LMC_SNR_Atlas}, but the SNR sizes in particular are not reliable and must be
revised. Sizes of SNRs with sharp outer boundaries are overestimated due to the large \textit{ROSAT} PSF
\citep[e.g. \ofo,][]{badenes07:outflows}, while sizes of diffuse SNRs are underestimated due to the low \textit{ROSAT}
effective area \citep[e.g. N23,][]{hughes06:N23}. We have searched the literature for more recent \textit{Chandra}
observations to constrain the LMC SNR sizes, and we have selected the eight smallest objects (sizes $<1.5$ arcmin, see
Table \ref{tab-1}).

The age estimates listed in Table \ref{tab-1} merit a few comments. For SNRs without SN or light echo information, ages
are calculated from the SNR size assuming a specific model for the SNR dynamics, which can introduce large
uncertainties. In particular, the standard dynamical models for young SNRs \citep[e.g.][]{truelove99:adiabatic-SNRs}
ignore the effect of cosmic ray acceleration at the forward shock. It is now widely accepted that energy losses due to
cosmic ray acceleration can affect the size of young SNRs in a noticeable way \citep{ellison04:hd+cr,warren05:Tycho},
which implies that calculations based on unmodified SNR dynamics can overestimate the age by as much as 20$\%$
\citep[see \S~5.2 in][for a discussion]{badenes07:outflows}.

\begin{figure*}
  
  \centering
  
  \includegraphics[scale=0.75,clip]{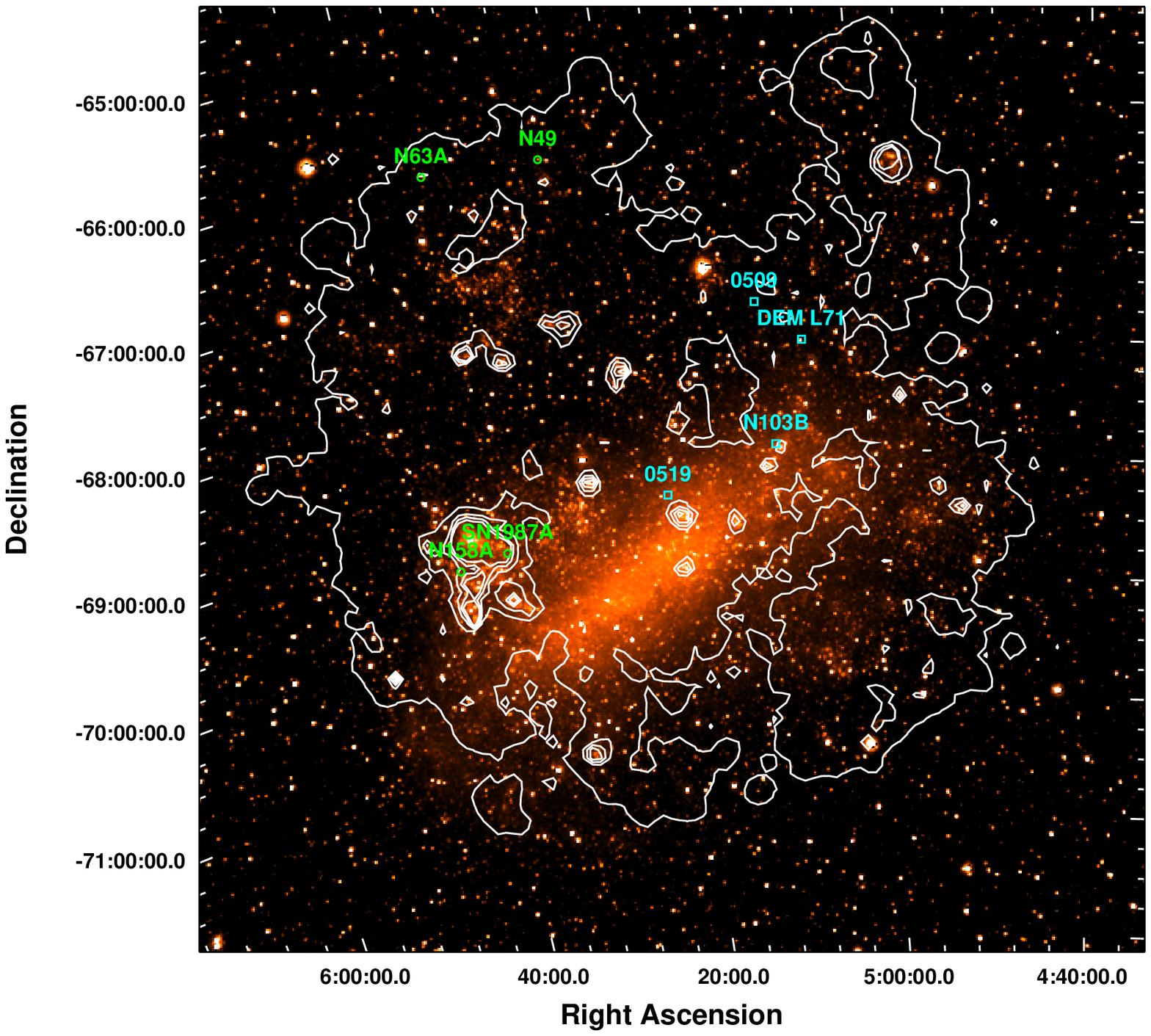}
  
  \caption{Map of the LMC in the SHASSA continuum band, indicating the positions of the eight target SNRs (core collapse
    SNRs with green circles, Type Ia SNRs with cyan squares). The overlaid contours are from the H$\alpha$ SHASSA image,
    which highlights the LMC disk, the W, S, and B spiral arms \citep{staveley03:LMC_HI_structure}, and the 30 Dor
    region around SN 1987A and N158A.\label{fig-1}}
  
\end{figure*}

In Figure \ref{fig-1}, we illustrate the location of our eight target SNRs within the large scale structure of the LMC
using the data from field 13 of the Southern H-Alpha Sky Survey Atlas \citep[SHASSA,][]{gaustad01:SHASSA}. Two SNRs, SN
1987A and N158A, are located in the 30 Dor region, the most prominent active star forming region in the LMC. Two more,
N49 and N63A, are in the northern part of the disk, embedded in the North Blue Arm discussed in HZ09 and
\citet{staveley03:LMC_HI_structure}. SNRs \ofn\ and N103B are in the outer parts of the LMC bar. The last two objects,
\ofo\ and DEM L71, are in rather inconspicuous parts of the LMC disk, in the area called the Northwest Void by
HZ09. More specific discussions about the location of each SNR will be given in \S~\ref{sec:SFH}.

\section{FROM SUPERNOVAE TO SUPERNOVA REMNANTS: CORE COLLAPSE VS. TYPE Ia}
\label{sec:SNtoSNR}

\subsection{Typing SNRs}

Typing SNRs as CC or Type Ia can be an uncertain and treacherous business. Both CC and Ia SNe deposit a similar amount
of kinetic energy ($\sim10^{51}$ erg) in the ambient medium (AM), which often makes it impossible to distinguish mature
CC from mature Ia SNRs based on their size or morphology alone. A much more reliable way to type SNRs is to examine the
evidence left behind by the explosion itself: X-ray spectrum from the SN ejecta and AM, SNR dynamics, and properties of
the compact object or pulsar wind nebula (PWN), if present. In general, this can only be done for relatively young
objects \citep[but see][]{hendrick03:LMC_SNRs,rakowski05:G337}. By using methods along these lines, we have been able to
determine the type of all the objects in our list with a high degree of confidence, and in some cases even the SN
subtypes within the broader CC and Ia categories. In this Section we will discuss the classification and SN subtypes of
our target SNRs, but before going into the details of each object, it is important to mention the work of
\citet{chu88:LMC_SNRs_environments}. These authors attempted to type \textit{all} the LMC SNRs known at the time by
noting the distance from each object to HII regions and OB associations. Although this is a very crude method, their
conclusions regarding the CC or Type Ia nature of our target SNRs coincide with ours, except in the case of SNR N103B,
which will be discussed in detail in \S~\ref{sec:SNtoSNR:IaSNRs} and \ref{sec:SFH:Ia}.

%



 



\subsection{Core Collapse SN Progenitors And Subtypes}
\label{sec:SNtoSNR:CC}

Our theoretical understanding of core collapse SNe is still incomplete \citep{janka07:CCSN_Review}. In particular, the
mapping between progenitor mass and CC SN subtype is uncertain, because key processes like stellar mass loss and binary
interactions are not well understood \citep{eldridge04:CCSN_Progenitors,eldridge08:CCSN_Progenitors_Binaries}. To set
the stage for further discussions, the stellar evolution models of \citet{eldridge04:CCSN_Progenitors} for single stars
of LMC metallicity ($Z=0.008$) predict that stars between $8$ and $30\,\mathrm{M_{\odot}}$ will explode as red
supergiants, retaining most of their H envelope and becoming Type IIP SNe, stars between $30$ and
$40\,\mathrm{M_{\odot}}$ will lose a large part of their envelopes and explode as Type IIL or Type IIb SNe, and stars
above $40\,\mathrm{M_{\odot}}$ will lose all their envelopes and become naked CC SNe of Types Ib and Ic. Within naked CC
SNe, there is some evidence that Type Ic SNe, which are linked to long duration gamma-ray bursts
\citep{galama98:SN-GRB,stanek03:SN_GRB} come from more massive stars than Type Ib SNe
\citep{anderson08:CCSN_progenitors_Halpha,kelly08:long_GRB-SNIc}. Stars that retain a massive H envelope but explode as
blue supergiants instead of red supergiants form a separate class, often referred to as SN 1987-like events.  The
lifetimes associated with these stellar masses range between 41 Myr for an isolated $8\,\mathrm{M_{\odot}}$ star and 5.4
Myr for an isolated $40\,\mathrm{M_{\odot}}$ star \citep[always taking the $Z=0.008$ models
from][]{eldridge04:CCSN_Progenitors}.  In principle, mass loss will be facilitated by binary interactions, leading to
fewer red supergiants and more Type Ib/c SNe in binary systems, but stellar evolution calculations that include these
effects are subject to an entirely different set of uncertainties \citep{eldridge08:CCSN_Progenitors_Binaries}.

From the point of view of the SNRs, the complex and turbulent structure of most young CC SNRs makes a quantitative
interpretation of the X-ray spectrum in terms of specific explosion models and progenitor scenarios very challenging
\citep[e.g. see][]{laming03:X-ray_knots_CasA,young06:CasA_Progenitor,park07:G292}. Many times, it is hard to infer the
SN subtype from the observational properties of the SNR, but the large intrinsic diversity of CC SNe as a class can
often be used to some advantage in SNR studies. \citet{chevalier05:young_cc_SNRs} argues that several aspects of SNR
evolution are expected to be very different depending on the subtype of the parent SN: mixing in the ejecta, fallback
onto the central compact object, expansion of the PWN, interaction with the CSM, and photoionization of the AM by shock
breakout radiation. Using arguments along these lines, \citet{chevalier03:CasA} inferred from the positions of the fluid
discontinuities, the presence of high velocity H, and the extent of the clumpy photoionized pre-SN wind in the Cas A SNR
that its progenitor must have been a Type IIn or Type IIb event. This `prediction of the past' was later confirmed by
the spectroscopy of the light echo of the Cas A SN \citep{krause08:CasA_light_echo}, which is very similar to the
spectrum of the Type IIb SN1993J. Although this agreement is certainly encouraging, we must insist that studies based on
SNRs are still a long way from providing a robust method of subtyping CC SNe - as an example, the Type IIn/IIb
classification of Cas A by \citet{chevalier03:CasA} was challenged by \citet{fesen06:CasA_highvelO}, who argued for a
Type Ib progenitor.

\subsection{Core Collapse SNRs}
\label{sec:SNtoSNR:CCSNRs}

In the following paragraphs, we examine each of the four target CC SNRs in more detail. For a summary, see Table
\ref{tab-2}.

\paragraph{SNR N49} This SNR harbors one of only two magnetars known outside the Milky Way: SGR 0526$-$66. In principle,
the presence of a compact object should immediately classify this object as a CC SNR, but the association between this
magnetar and the SNR has been controversial \citep{gaensler01:AXP_SGR_SNR}.  Even disregarding the compact object, the
ejecta emission shows significantly enhanced abundances from O and Si, but a comparatively small amount of Fe
\citep{park03:N49}, and the SNR is located within the OB association LH53 \citep{chu88:LMC_SNRs_environments}. Taken
together, these arguments lend strong support to a CC origin. The complex filamentary structure of the shocked AM
suggests that dense material surrounded the SN at the time of the explosion \citep{bilikova07:N49}, which favors a
progenitor with a slow wind, maybe a Type IIP SN. Unfortunately, this is just an educated guess, because the complex
multi-phase X-ray emission of the SNR and the poorly known age make the interpretation of the observations very
challenging.

\paragraph{SNR N63A} This SNR has no detected compact object, although the upper limits do not exclude the presence of a
low-activity PWN \citep{warren03:N63A}. It is embedded in the large HII region N63, and it also appears to be located
within an OB association \citep[NGC 2030,][]{chu88:LMC_SNRs_environments}, making a CC type very likely.  The
size, morphology, and X-ray spectrum show evident signs of a complex interaction with a highly structured AM, and they
seem to indicate that it is expanding into a large cavity \citep{hughes98:LMC_SNRs_ASCA}, which suggests a massive
progenitor with a fast wind, maybe a Type Ib/c SN. However, there is no additional evidence to support this conclusion
because the X-ray emission, which is dominated by the shocked AM, reveals very little about the properties of the SN
ejecta \citep{warren03:N63A}.

\paragraph{SN 1987A} The classification and subtype of SN 1987A are obvious from the SN spectroscopy. The vast amount of
information available on this object is summarized in \citet{mccray07:SN1987A_20year} and other publications in the same
volume. For our purposes, it suffices to mention that the progenitor of this SN is known to have been a blue supergiant
star, Sk $-69^{\circ} 202$, whose initial mass has been estimated at $\sim20\, \mathrm{M_{\odot}}$
\citep{arnett91:1987A}, and might have been part of a close binary system \citep{podsiadlowski90:merger_87A}.

\paragraph{SNR N158A} This object harbors a well-observed PWN that types it as a CC SNR and constrains its age to be
$\sim$800 yr \citep{kirshner89:0540,chevalier05:young_cc_SNRs}. The SN subtype classification has been rather
controversial. The SNR dynamics indicate that the shock wave is moving into dense, clumpy CSM similar to what can be
found around a massive Wolf-Rayet star, and the presence of strong O and S lines in the X-ray spectrum of the innermost
ejecta reveal that at least some of the heavy elements in the ejecta have not fallen back onto the central neutron star
\citep{chevalier05:young_cc_SNRs}. This would favor a massive Type Ib/c progenitor, but both the detection of H in the
PWN filaments \citep{serafimovich05:SNR0540-69.3} and a recent re-analysis of the ejecta emission
\citep{williams08:0540} seem to indicate that the progenitor might have been in the $20-25\,\mathrm{M_{\odot}}$ range,
implying a Type IIP explosion \citep[for more detailed discussions,
see][]{chevalier06:progenitor_to_afterlife,williams08:0540}.

\subsection{Type Ia SN Progenitors and Subtypes} 
\label{sec:SNtoSNR:Ia}

Our current understanding of Type Ia SN progenitors is still extremely sketchy
\citep{maoz08:fraction_intermediate_stars_Ia_progenitors}, but several interesting trends have been inferred from the
bulk properties the host galaxies. Any theoretical model for Type Ia progenitors must account for the fact that Type Ia
SNe explode in elliptical galaxies with very little star formation (SF), but at the same time the rate of Ia events in
star forming galaxies appears to scale with the specific star formation rate
\citep{mannucci06:two_progenitor_populations_SNIa}. Moreover, Type Ia SNe exploding in elliptical galaxies are on
average dimmer that those exploding in star forming galaxies
\citep{hamuy96:Ia-absoluteM,hamuy00:Environment_effects_SNIa}. \citet{scannapieco05:A+B_models} and
\citet{mannucci06:two_progenitor_populations_SNIa} used these observational facts to postulate \textit{two} populations
of Type Ia SN progenitors: a `prompt' population with short delay times (of the order of a few hundred Myr), associated
with recent SF and leading to somewhat brighter SN Ia, and a `delayed' population with longer delay times (of the order
of Gyr), not associated with recent SF and leading to somewhat dimmer SN Ia. In this two-component model, the specific
Type Ia SN rate in a given galaxy is expressed as $SNR_{Ia}=A M_{*} + B \dot{M_{*}}$, with $M_{*}$ being the total
stellar mass in the galaxy and $\dot{M_{*}}$ the specific star formation rate. It is important to stress that the
observed rates do not \textit{require} the existence of two components - in some theoretical scenarios, Type Ia SNe from
a single progenitor channel can explode with both very short and very long delay times
\citep{greggio08:SN_Ia_rates}. Nevertheless, there have been several attempts to associate the prompt and delayed
progenitor populations to the two leading theoretical scenarios for Type Ia SNe: single degenerate (SD) systems, in
which the WD accretes material from a non-degenerate companion and double degenerate (DD) systems, in which the WD
accretes material from another WD. So far, none of these attempts has succeeded \citep[see
f.i.][]{forster06:SNIa_progenitor_delays,greggio08:SN_Ia_rates}, and the identity of Type Ia SN progenitors remains a
mystery.

Despite all the uncertainties regarding their progenitors, Type Ia SN explosions as a class are much more homogeneous
and have less intrinsic dispersion than CC SNe. In particular, there is a simple relationship between the structure of
the ejecta and the peak brightness of the SN that is well reproduced by one-dimensional delayed detonation (DDT)
explosion models \citep{mazzali07:zorro}. This allows to map the vast majority of SN Ia onto a sequence of bright to dim
events based on the amount of $^{56}$Ni that they synthesize. Generally speaking, Type Ia SNRs are also much less
turbulent than CC SNRs, and most of them seem to be interacting with a relatively unmodified AM
\citep{badenes07:outflows}, although there are exceptions like the Kepler SNR \citep{reynolds07:kepler} and SNR N103B
\citep[][, see also the discussion in \S~\ref{sec:Conc-Ia}]{lewis03:N103B}. Thanks to this set of circumstances, it is
generally easier to interpret the X-ray emission of Type Ia SNRs quantitatively in terms of specific explosion models,
provided that the dynamic evolution of the SNR and the nonequilibrium processes in the shocked plasma are properly taken
into account \citep{badenes03:xray,badenes04:PhD,badenes05:xray}. It is also possible to estimate the brightness of the
parent event from the mass of $^{56}$Ni synthesized by the preferred DDT explosion model, as shown by
\citet{badenes06:tycho} in the case of the Tycho SNR.

\subsection{Type Ia SNRs} 
\label{sec:SNtoSNR:IaSNRs}

These four objects were classified as Type Ia SNRs by \citet{hughes95:typing_SN_from_SNR} based on their lack of compact
object or PWN and the general properties of their X-ray emission, which is dominated by Fe lines and has only weak or
absent lines from O. In order to confirm these classifications and derive the SN subtype, it is necessary to perform an
in-depth analysis of the ejecta emission, as done by \citep{badenes08:0509} for SNR \ofo. The X-ray emission of the
other three objects will be the subject of a forthcoming publication \citep[][henceforth BH09]{badenes09:IaSNRs_LMC},
but the main results of that analysis are presented in the following paragraphs, and summarized in Table \ref{tab-3}.

\paragraph{SNR DEM L71} This is the oldest object in our Type Ia SNR list, and the only one without a light echo age
estimate. \citet{ghavamian03:DEML71} determine an age of $4360 \pm 290$ from the SNR dynamics, and yet the X-ray
spectrum appears dominated by shocked Fe from the SN ejecta, specially in the center \citep{hughes03:DEML71}. The old
age of this SNR makes the analysis of the ejecta emission somewhat challenging, but BH09 find that it can be reproduced
by DDT models for normal Type Ia SNe.

\paragraph{SNR N103B} This object was initially classified as a CC SNR by \citet{chu88:LMC_SNRs_environments} based on
its location at the edge of the HII region DEM 84 and 40 pc away from the OB association NGC 1850. However, the X-ray
spectrum is strongly suggestive of a Type Ia origin \citep{hughes95:typing_SN_from_SNR,lewis03:N103B}. The SNR is also
remarkable in that it shows a strong east-west asymmetry \citep{lewis03:N103B}, which has been interpreted as a sign of
some kind of CSM interaction, mainly by analogy to the Kepler SNR. This asymmetry also makes the ejecta analysis
challenging, but BH09 find a relatively good match to the spectrum using DDT models for moderately bright Type Ia SNe
with a SNR age close to the 800 yr estimated from the light echo by \citet{rest05:LMC_light_echoes}.

\paragraph{SNR \ofo} Next to SN1987A, this SNR has the most secure subtype classification in the LMC. In 2008, two teams
analyzed independently the optical spectrum of the SN light echo \citep{rest08:0509} and the X-ray emission and dynamics
of the SNR \citep{badenes08:0509}, and came to the same conclusion: SNR \ofo\ was originated $\sim400$ yr ago by an
exceptionally bright Type Ia SN that synthesized $\sim1\,\mathrm{M_{\odot}}$ of $^{56}$Ni. This agreement is a very
important validation of the modeling techniques introduced in \citet{badenes03:xray} that BH09 apply to the other three
target Type Ia SNRs, and in particular of the capability of the models to recover the SN subtype\footnote{The recent
  spectroscopic analysis of the light echo from Tycho's SN by \citet{krause08:tycho} also confirms the previous result
  by \citet{badenes06:tycho} based on the X-ray emission from the SNR that the SN was of normal brightness, not
  overluminous or underluminous.}.

\paragraph{SNR \ofn} The final object in our Type Ia SNR list has an estimated age of $600 \pm 200$ yr from its light
echo \citep{rest05:LMC_light_echoes}. Its X-ray emission is well reproduced by a moderately bright Type Ia SN model that
synthesizes $0.8\,\mathrm{M_{\odot}}$ of $^{56}$Ni (BH09).

\section{OVERVIEW OF THE STAR FORMATION HISTORY MAP OF THE LMC}
\label{sec:Map}

The SFH map that we use in the present work is described in full detail in HZ09. The map was elaborated using four band
(\textit{U}, \textit{B}, \textit{V}, and \textit{I}) photometry from the Magellanic Clouds Photometric Survey
\citep{zaritsky04:MCPS}, which has a limiting magnitude between $20$ and $21$ in \textit{V}, depending on the local
degree of crowding in the images. Data from more than 20 million stars was assembled to produce color-magnitude diagrams
in 500 $24'\times24'$ cells encompassing the central $8^\circ\times8^\circ$ of the LMC (see Figure 4 in HZ09), and then
the StarFISH code \citep{harris01:StarFISH} was applied to derive the local SFH for each cell. Cells with enough stars
in them, like the eight cells that contain our target SNRs, were further subdivided into four $12'\times12'$
subcells. The SFH of each cell is given at thirteen lookback times between $Log(t)=6.8$ (6.3 Myr) and $Log(t)=10.25$
(17.8 Gyr), and it is broken into four metallicity bins: $Z=0.008,\, 0.004,\, 0.0025,$ and $0.001$. 

\begin{figure*}

  \centering
 
  \includegraphics[scale=0.75,angle=90]{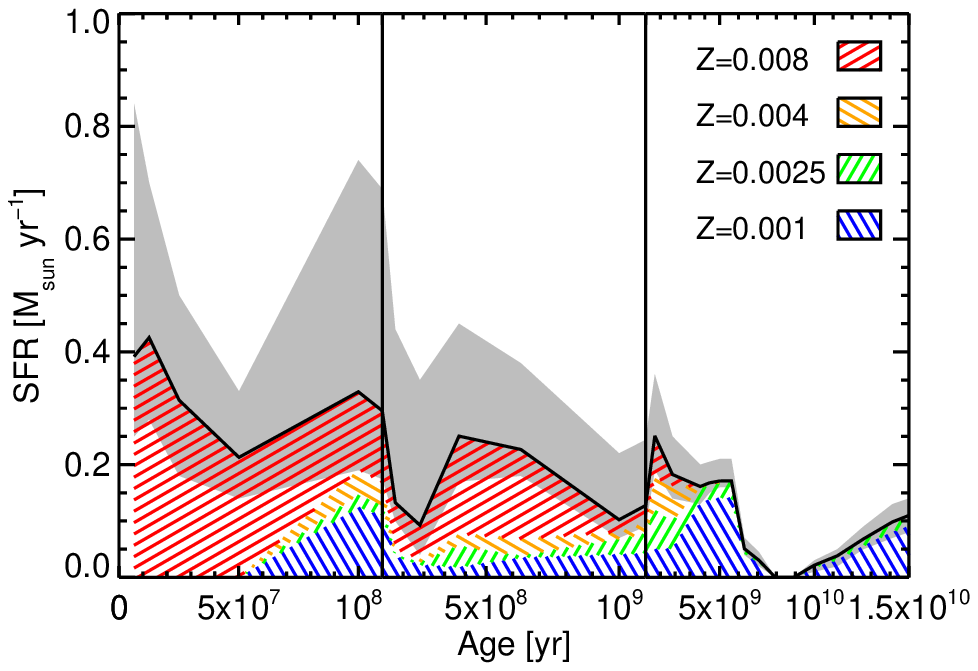}
  \includegraphics[scale=0.75,angle=90]{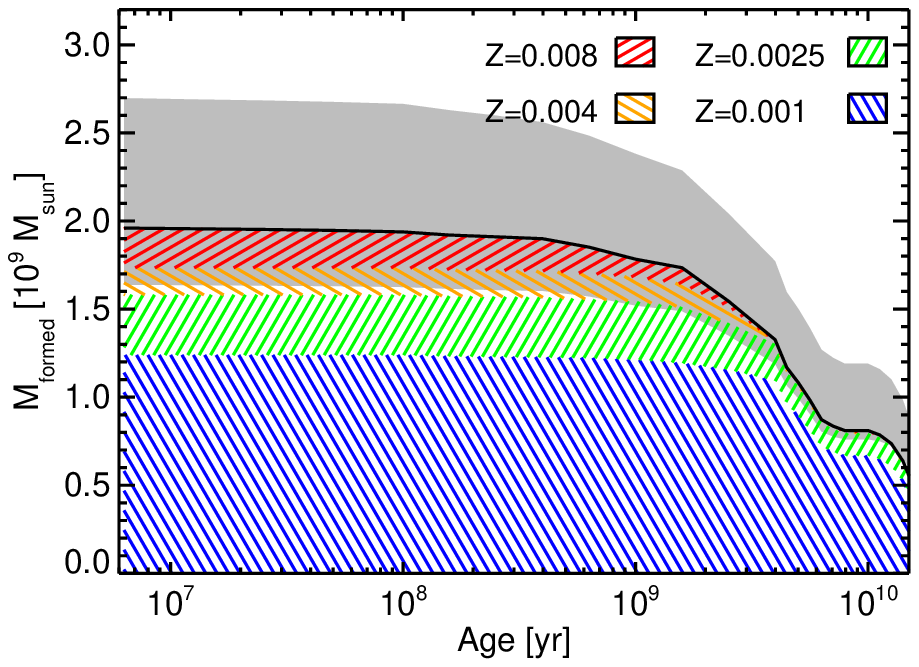}

  \caption{Total SFH of the LMC, broken into four metallicity bins. Left panel: star formation rate (SFR) as a function
    of lookback time. This plot is drawn following the style of HZ09 in three linear-linear panels that highlight the
    structure of the SFH at short (0 to 110 Myr), medium (110 Myr to 1.1 Gyr) and large (1.1 Gyr and beyond) lookback
    times. The gray area represents the error on the total SFR. Right panel: integrated SFR displaying the cumulative
    stellar mass formed in the LMC as a function of lookback time. Adapted from Figure 11 in HZ09. \label{fig-2}}

\end{figure*}

For reference in further discussions, we reproduce the SFH of the entire LMC from HZ09 in Figure \ref{fig-2}. The error
bars on the total SFH represented with the gray shaded area are dominated by crowding effects (see \S~3.3 in HZ09).
Although no metallicities were fitted for ages below 50 Myr, the plots display the canonical LMC metallicity ($Z=0.008$)
in the most recent age bins, which is reasonable in view of the high degree of homogeneity in the metallicity of the ISM
and the young stars in the LMC
\citep{pagel78:Chemical_Composition_HII_MCs,russell90:LMC_Abundances_II,korn02:LMC_NGC2004,hunter07:MC_BStars}. A
detailed discussion of the SFH of the LMC and its interpretation in the context of the LMC's past history can be found
in \S~5 of HZ09. For our purposes, it suffices to note that, after an initial episode of SF in the distant past, the LMC
went into a quiescent period that lasted until 5 Gyr ago, and since then it has been forming stars at an average rate of
$0.2\,\mathrm{M_{\odot}yr^{-1}}$, with episodes of enhanced SF at 2 Gyr, 500 Myr, 100 Myr, and 12 Myr. From Figure
\ref{fig-2}, it is obvious that the vast majority of the stars in the LMC have ages above 1 Gyr.  Most of these old
stars have metallicities of one tenth solar or lower.

\section{ON THE RELEVANCE OF THE LOCAL STELLAR POPULATIONS TO SUPERNOVA PROGENITORS}
\label{sec:Relevance}

During the lifetime of a galaxy, several processes naturally mix the stellar populations. These include both internal
processes like the `churning' of the disk by spiral arms \citep{sellwood02:radial_mixing_disks} and external processes
like tidal interactions and mergers \citep{mihos94:pop_gradients_mergers}. In this context, the properties of the
stellar population (and hence the SFH) in the neighborhood of a young SNR will only be representative of the SN
progenitor up to a certain lookback time, $t_{lb}$. In principle, $t_{lb}$ can be calculated for each location within a
galactic disk provided there is a viable dynamic model that includes all the relevant processes. Unfortunately, no such
model exists for the LMC, despite the wealth of observational information available. The LMC disk is warped
\citep{nikolaev04:LMC_Disk} and might also be flared \citep{subramanian08:LMC_SMC_Depth}, and it has a rich history of
tidal interactions with the SMC and (maybe) the Milky Way, which may have important effects on the stellar dynamics
\citep[see][]{olsen07:tidal_effects_LMC,besla07:MC_passage}. The vestigial arms seen in HI
\citep{staveley03:LMC_HI_structure} are probably originated by these tidal interactions, but the details of this process
are not well understood \citep{besla07:MC_passage}. Even the nature of the most prominent feature in the disk - the LMC
bar - and its role in the dynamics of the galaxy are unclear \citep{zaritsky04:LMC_bar}.

Without a reliable way to calculate $t_{lb}$ for each of the subcells that contain our target SNRs, all we can do is
estimate the relevant timescales for a number of different processes. The physical size of the subcells in the HZ09 map
is $350 \times 350$ pc \citep[assuming $D=50$ kpc,][]{alves04:LMC_Distance}, and the velocity dispersions for the young
disk and old disk populations determined by \citet{graff00:LMC_velocity_structure} are $8$ and
$22\,\mathrm{km\,s^{-1}}$, respectively. Thus, the length of time that it would take an average star of the young (old)
disk to drift from one subcell to the next in the absence of restoring forces, $t_{d}$, is 43 (16) Myr. These timescales
are not relevant for the progenitors of CC SNe, which should belong to the young disk, but they will be very important
for SN Ia progenitors, which could be quite older. In any case, $t_{lb}$ should be much larger than $t_{d}$, because (a)
some regions of the LMC are more homogeneous than others, which means that stars have to drift over larger distances in
order to find substantially different stellar populations, and (b) there are restoring forces like gravity that maintain
the structural integrity of the disk and act to limit stellar drift.

\begin{figure*}

  \centering
 
  \includegraphics[scale=0.75,angle=90]{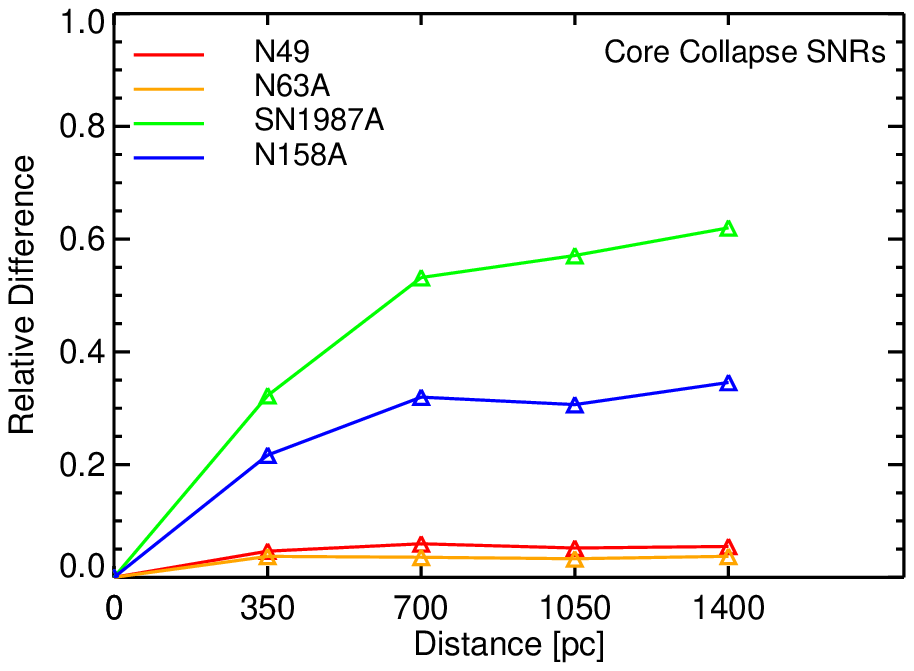}
  \includegraphics[scale=0.75,angle=90]{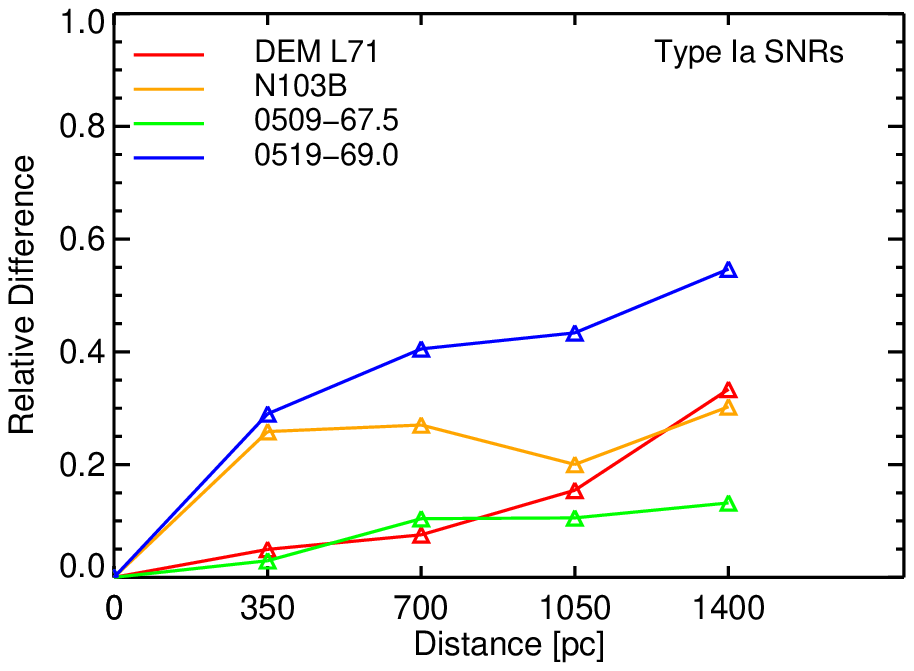}

  \caption{Relative difference in the stellar populations as a function of distance from the center of the subcells
    containing our eight target SNRs. Left panel: core collapse SNRs; right panel: Type Ia SNRs. \label{fig-3}}

\end{figure*}

We have quantified the spatial homogeneity of the stellar populations around our target SNRs in Figure \ref{fig-3}. We
plot the absolute value of the relative differences in the stellar populations as a function of the distance from the
center of each of the eight subcells that contain the CC and Ia SNRs in our list. To calculate the relative differences,
we have integrated the SFH in each neighboring subcell, taking all the time bins were the differences between the
neighbor and the SNR subcell were statistically significant (i.e., the error bars did not overlap) up to a lookback time
of 1.1 Gyr, and then divided by the total number of stars formed in the SNR subcell. At each distance, the relative
difference is the mean of the relative differences between the SNR subcell and all the subcells at that distance. Figure
\ref{fig-3} shows that some of our target SNRs are in remarkably homogeneous regions of the LMC disk. These include the
CC SNRs N49 and N63A in the Blue Arm and the Type Ia SNR \ofo\ in the Northwest Void, with average relative differences
in the stellar populations below $15\%$ within 1400 pc of the central subcell. This distance translates into $t_{d}$
values of 215 and 80 Myr for young and old disk stars in the absence of restoring forces.

The effect of restoring forces on the value of $t_{lb}$ is more difficult to estimate. If no chaotic processes
intervene, neighboring stars will tend to move together through the disk, which explains why some LMC structures like
the bar and the Northwest Arm show up in the HZ09 map with lookback times as large as 1 Gyr (see their Figure 8). This
long survival time is not restricted to large structures - in the Solar neighborhood, there is evidence that several
groups of old stars (2 to 8 Gyr) are moving together through the disk of the Milky Way
\citep{dehnen98:Nearby_Stars_Velocity_Space}. But smaller structures like young stellar clusters seem to disappear on
timescales of the order of $180$ Myr in the LMC \citep{bastian08:LMC_clusters}, which is roughly equivalent to the
dynamic crossing time of the LMC disk. We will adopt this value as a figure of merit for $t_{lb}$ in a single subcell of
the SFH maps. Since the result of \citet{bastian08:LMC_clusters} applies to young stars, this implies that restoring
forces increase the value of $t_{d}$ by at least a factor $\sim4$, but we stress that this is just a very rough
estimate.

We conclude that the relevance of the local SFHs for Type Ia SN progenitors will depend on both the homogeneity of the
stellar populations around each subcell (Figure \ref{fig-3}) and the age of the progenitors. If the lifetime of Type Ia
progenitors in the prompt channel is as short as the 180 Myr claimed by \citet{aubourg07:massive_stars_SNIa}, it should
be possible to use the local SFHs of Type Ia SNRs in the LMC to explore their properties. For progenitors with longer
lifetimes, the stellar context of each SNR should be taken into account. Objects like SNR \ofo\ might allow exploration
of timescales up to several hundred Myr, but SNRs like \ofn\ probably will not.

\section{STAR FORMATION HISTORIES AROUND THE TARGET SNRs}
\label{sec:SFH}

The local SFHs in the subcells containing the eight SNRs in our list are plotted in Figures \ref{fig-4} and
\ref{fig-6}. The lifetime of an isolated $8\,\mathrm{M_{\odot}}$ star with $Z=0.008$ from
\citet{eldridge04:CCSN_Progenitors} has been indicated by a dashed vertical line in all the plots for illustrative
purposes. For simplicity, we have collapsed all the SFH bins at ages above 2 Gyr into a single bin at 10 Gyr. Several
interesting average quantities can be calculated from the local SFHs. We have listed two such quantities in Tables
\ref{tab-2} and \ref{tab-3}: the average metallicity of all the stars formed in the subcell, $\bar{Z_{*}}$ and their
average age $\bar{t_{*}}$. These averages are always dominated by the large population of old stars in each subcell, and
therefore they are irrelevant for the properties of CC SN progenitors - the values in Table \ref{tab-2} are merely
provided for comparison with the values in Table \ref{tab-3} (see discussion in \S~\ref{sec:Disc}). The average values
for the entire LMC are $\bar{Z_{*}}=0.0023$ and $\bar{t_{*}}=8.1$ Gyr.

\subsection{Core Collapse SNRs}
\label{sec:SFH:CC}

\begin{figure*}

  \centering

  \includegraphics[scale=0.75,angle=90]{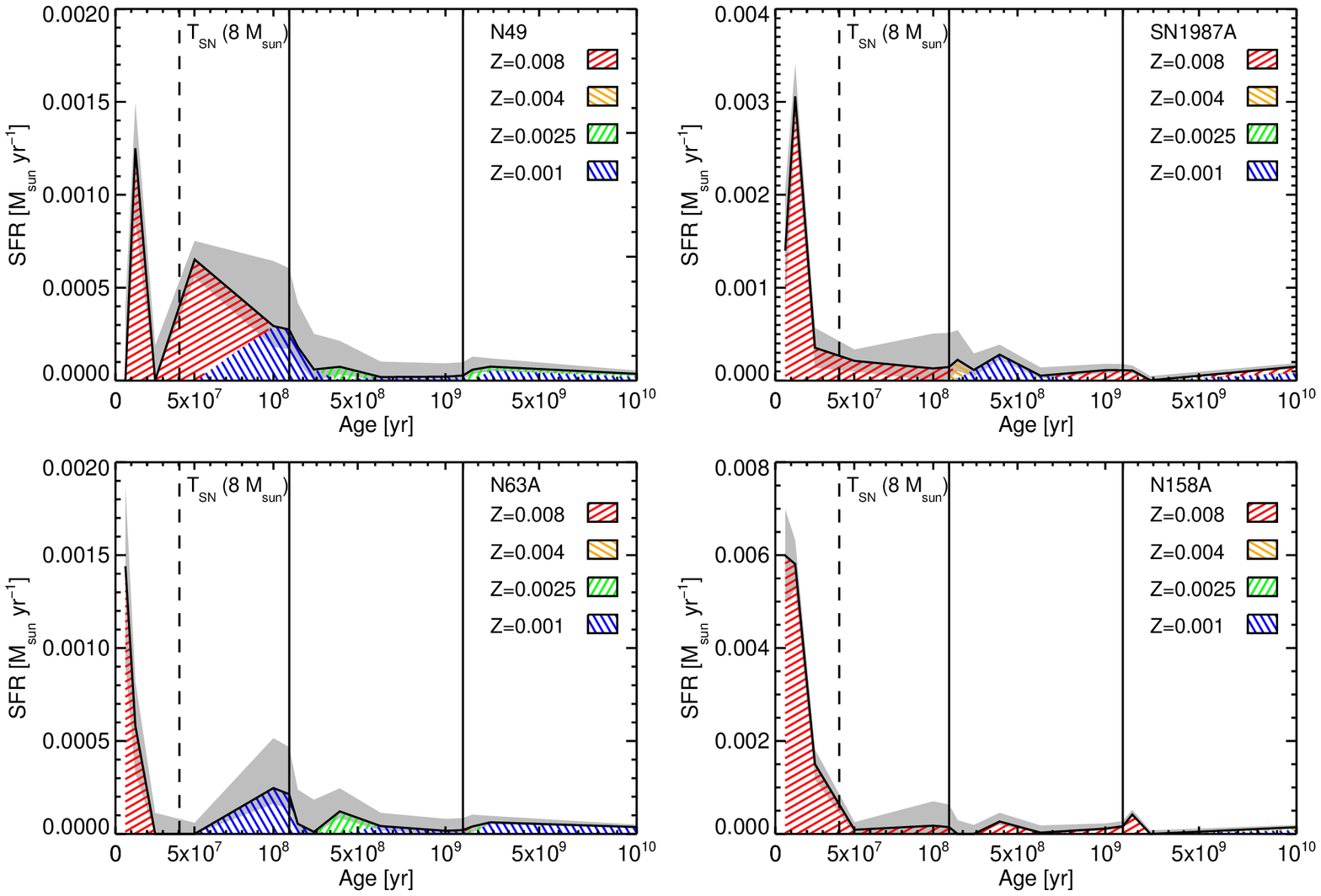}  

  \caption{Local SFHs around the four target core collapse SNRs. See Figure \ref{fig-2} for an explanation of the
    plots. \label{fig-4}}

\end{figure*}

\begin{figure}

  \centering
  
  \includegraphics[scale=0.75,angle=90]{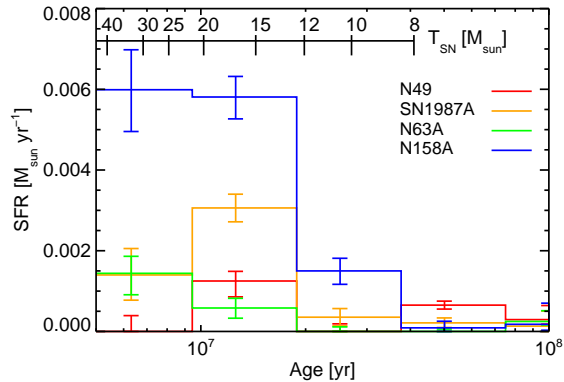}  

  \caption{Detail of the recent SFH around the four target CC SNRs. The lifetimes of isolated stars of LMC metallicity
    with different masses are displayed on the ruler on top of the plot. \label{fig-5}}

\end{figure}

\begin{figure*}

  \centering
  
  \includegraphics[scale=0.75,angle=90]{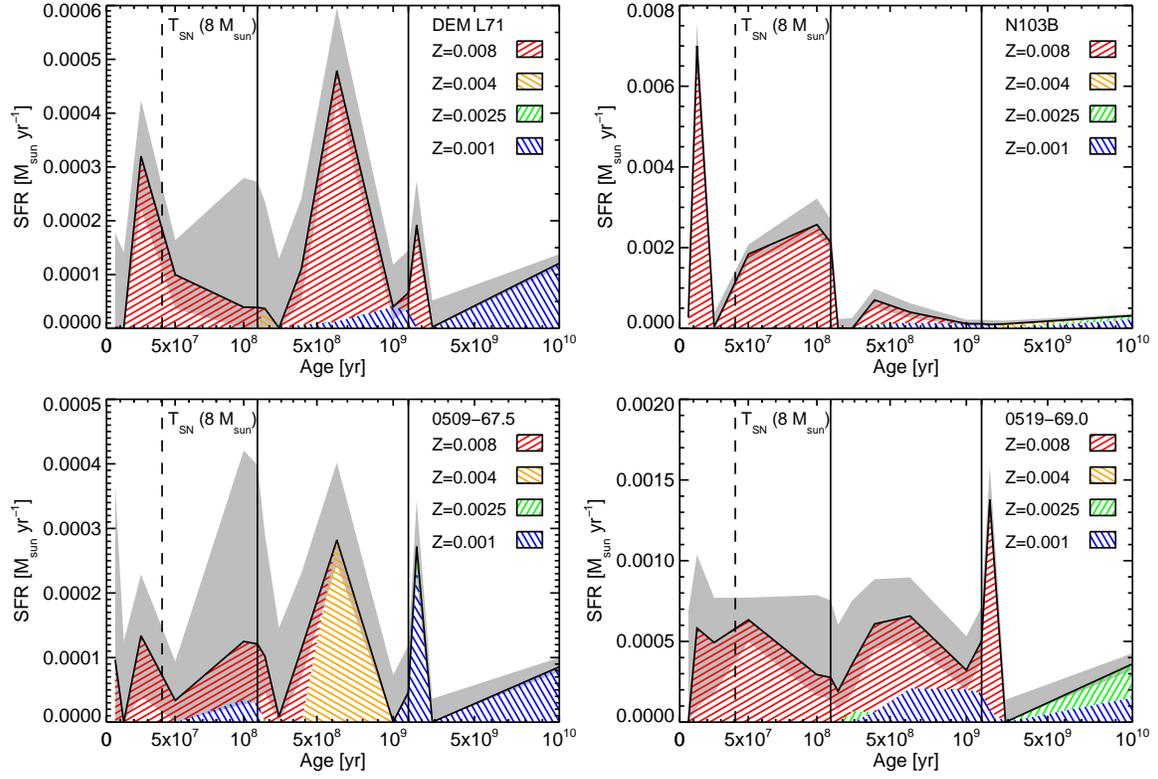}  

  \caption{Local SFHs around the four target Ia SNRs. See Figure \ref{fig-2} for an explanation of the
    plots. \label{fig-6}}

  \vspace{1cm}

\end{figure*}

The most salient feature of the SFHs around the four CC SNRs (Figure \ref{fig-4}) is that they are strongly dominated by
intense bursts of star formation in the recent past ($t<40$ Myr). This is of course expected, and in the cases where the
SNRs have been typed for their close proximity to young stellar clusters (notably, SNR N63A), it does not reveal any new
information. However, the timing of these bursts and their intensity will determine the properties of the population of
massive stars that can be found at each location, and hence the likelihood of each CC SN subtype. To highlight these
aspects, we display the most recent bins of the SFHs associated with the CC SNRs in greater detail in Figure
\ref{fig-5}, alongside the lifetimes of isolated massive stars with $Z=0.008$ from \citet{eldridge04:CCSN_Progenitors}
\footnote{Other grids of stellar models \citep[e.g.][]{maeder89:Stellar_Models,girardi00:Low_Intermediate_Mass_Stars}
  can give slightly different values for the lifetimes of isolated stars of LMC metallicity. In general, these
  differences are not large enough to have an impact on our work.}. We have convolved the three most recent SFH bins
with a standard Salpeter IMF to calculate the fraction of massive stars that are exploding now as CC SNe ($f_{CCSN}$)
from progenitors in three mass intervals: $8$ to $12.5\,\mathrm{M_{\odot}}$, $12.5$ to $21.5\,\mathrm{M_{\odot}}$ and
above $21.5\, \mathrm{M_{\odot}}$. We list these fractions in Table \ref{tab-2} for each of the CC SNRs. The interval
cuts are the stellar masses whose lifetimes correspond to the upper edges of the first and second age bins in the SFHs:
9.4 and 18.9 Myr. We remind the reader that these values of $f_{CCSN}$ are calculated using isolated star models that do
not take into account the potentially large effects of binarity on stellar evolution. An entirely different but also
potentially serious problem comes from the fact that massive stars are notoriously difficult to study using photometry
alone \citep{massey95:massive_stars_field}.  Because StarFISH uses \textit{all} the stars (not just the massive ones) in
each subcell to calculate the SFR at each age, the values of $f_{CCSN}$ might not be severely affected by this, but to this
date there has been no systematic calibration of the StarFISH results for young stellar populations. To reflect these
and other caveats, we do not list error bars on the $f_{CCSN}$ values, which should only be regarded as approximate.

\paragraph{SNR N49} The integrated SFH for the North Blue Arm region that contains SNR N49 and SNR N63A is dominated by
a coherent episode of low-metallicity star formation 100 Myr ago (see \S~5.1.3 and Figure 16 in HZ09), which is apparent
in the corresponding panels of Figure \ref{fig-4}. This is the reason why the values of $\bar{Z_{*}}$ and $\bar{t_{*}}$
for SNRs N49 and N63A are lower than those of the other SNRs. Many parts of the North Blue Arm have also had noticeable
star formation activity in the last 40 Myr, although not as intense as in the more prominent star forming regions of the
LMC, 30 Dor and Constellation III. SNR N49 is in one such region, which had a moderately intense SF burst 12 Myr ago,
but very little SF activity in the most recent bin centered at 6.3 Myr (see Figure \ref{fig-5}). From these properties
of the SFH, the expectation is that the majority of the CC SN progenitors in this subcell should be stars between $12.5$
and $21.5\,\mathrm{M_{\odot}}$ (see Table \ref{tab-2}). Even taking the upper limit of the SFR in the most recent bin
and the lower limit on the bin at 12 Myr, the fraction of CC SN progenitors with masses above $21.5\,\mathrm{M_{\odot}}$
remains below $1\%$, with all the caveats associated to the calculated values of $f_{CCSN}$. This is in good agreement
with the properties of the SNR discussed in \S~\ref{sec:SNtoSNR:CCSNRs}, and it has interesting implications for the
association of SNR N49 with SGR 0526$-$66. Magnetars are thought to be originated by very massive
($>30\,\mathrm{M_{\odot}}$) stars
\citep{gaensler05:magnetars_massive_stars,figer05:Massive_stars_SGR1806-20,muno08:magnetar_search}, but such stars
appear to be very scarce around SNR N49. One possibility is that the magnetar was formed elsewhere and the association
is coincidental. \citet{gaensler01:AXP_SGR_SNR} examined this issue in detail, and came to the conclusion that the link
between SNR N49 and SGR 0526$-$66 is considerably less convincing than those of other magnetars in SNRs. More recently,
\citet{klose04:NIR_survey_N49_SGR0526-66} performed a NIR survey of the area around SNR N49 and identified a young
stellar cluster (SL 463) at a projected distance of $\sim30$ pc northeast of SGR 0526$-$66 that could have been the
birthplace of the magnetar. This cluster does fall partially on a neighboring subcell of the HZ09 map with intense SF at
6.3 Myr, consistent with the 5 to 20 Myr age estimates for SL 463 by \citeauthor{klose04:NIR_survey_N49_SGR0526-66}, and
much more promising as a birthplace of massive CC progenitors ($36\%$ above $21.5\,\mathrm{M_{\odot}}$). As pointed out
by \citeauthor{klose04:NIR_survey_N49_SGR0526-66}, if this hypothesis is true the magnetar must have been ejected from
its birthplace with a certain velocity, and should have a measurable proper motion (see their \S~3.2). Another
possibility is that the association between SNR N49 and SGR 0526$-$66 is indeed real, but not all magnetars have stellar
progenitors more massive than $30\,\mathrm{M_{\odot}}$.

\paragraph{SN 1987A} The SFH associated with this SNR is of particular interest because, together with the known mass of
the progenitor, $\sim20\,\mathrm{M_{\odot}}$, it can provide some test of the robustness of our SFH approach to CC SN
progenitors. \citet{panagia00:Stars_around_87A} conducted a detailed study of the immediate neighborhood of SN1987A
within the 30 Dor region using \textit{HST} data, and found a loose young cluster with an age of $12\pm2$ Myr, which
they identified as the birthplace of SN1987A's progenitor. The local SFH drawn from the HZ09 map is indeed dominated by
an intense SF episode 12 Myr ago, in good agreement with the results of \citet{panagia00:Stars_around_87A}. From the
SFH, we expect $56\%$ of the CC SN progenitors in this region to be stars between $12.5$ and $21.5\,\mathrm{M_{\odot}}$
(see Figure \ref{fig-5}). This agreement is encouraging, and indicates that some information about the progenitor mass
of CC SNe can be recovered from the SFH maps of HZ09.

\paragraph{SNR N63A} Like SNR N49, SNR N63A is located in a part of the Blue Arm with SF activity in the recent
past. The SFH in this subcell, however, is different from that around SNR N49 in that it is dominated by the most recent
bin centered at 6.3 Myr. As a result, $70\%$ of the CC SN progenitors in this subcell are expected to be stars more
massive than $21.5\,\mathrm{M_{\odot}}$. With a Salpeter IMF, roughly $40\%$ of these stars will be in turn more massive
than $40\,\mathrm{M_{\odot}}$, which makes a Type Ib/c origin for SNR N63 plausible, as suggested by the properties of
the SNR. If this were true, SNR 63A would be one of the youngest nearby Type Ib/c SNRs, and a closer examination of this
object to locate the elusive compact object and study the ejecta emission in greater detail would be of the highest
interest.

\paragraph{SNR N158A} This object is also located in 30 Dor, but in a region with even more vigorous recent SF than the
neighborhood of SN 1987A. According to our estimates, $24\%$ of the CC SN progenitors in this subcell should be stars
more massive than $21.5\,\mathrm{M_{\odot}}$. Again, this is compatible with the relatively massive progenitor suggested
by the SNR properties (see \S~\ref{sec:SNtoSNR:CC}), but unfortunately the temporal resolution of the SFH does not allow
us to discriminate between a Type IIP progenitor with $20\, -\, 25\,\mathrm{M_{\odot}}$ \citep{williams08:0540} and a
Type Ib/c progenitor with $>40\,\mathrm{M_{\odot}}$ \citep{chevalier05:young_cc_SNRs}. From the point of view of the
stellar population around SNR 158A, both hypotheses are equally plausible.

\subsection{Type Ia SNRs}
\label{sec:SFH:Ia}

\begin{figure*}

  \centering
  
  \includegraphics[scale=0.75,angle=90]{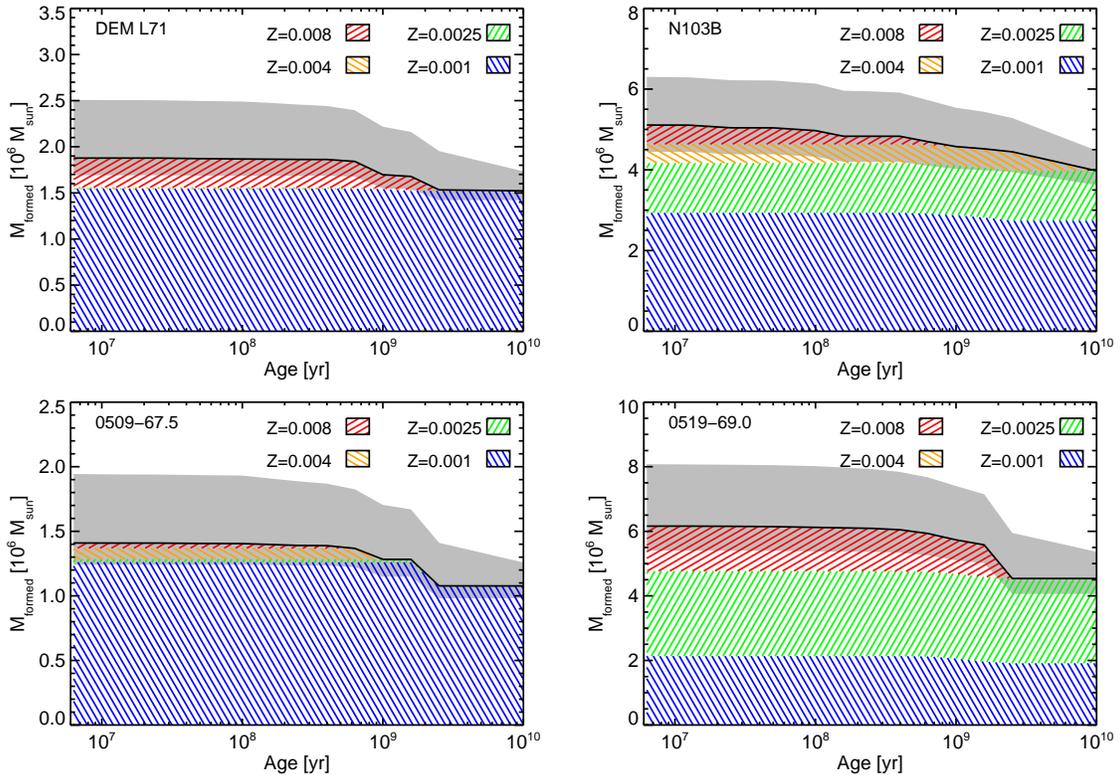}  

  \caption{Integrated SFHs around the four target Ia SNRs. See Figure \ref{fig-2} for an explanation of the
    plots. \label{fig-7}}

\end{figure*}

The local SFHs around the four Type Ia SNRs are displayed in Figure \ref{fig-6}. Since the age of the progenitors is not
known \textit{a priori}, we also display the integrated SFHs in Figure \ref{fig-7} to provide a more intuitive picture
of the makeup of the stellar populations in these locations and how they have evolved with time. With one exception (SNR
N103B, see below), these SFHs are very different from those associated with the CC SNRs. The SFHs of SNRs DEM L71, \ofo,
and \ofn\ show very little (but nonzero) activity in the last 200 Myr, resulting in old and metal poor stellar
populations, typical of the regions of the LMC without ongoing star formation. The SFHs of these three SNRs are also
punctuated by bursts of SF at 600 Myr and 2 Gyr whose prominence varies from object to object, but this is probably more
related to the global properties of the LMC (see Figure \ref{fig-2}) than to the Type Ia SN progenitors.

Given the coincidence of the LMC disk crossing time and the upper limit for the delay time of short-lived SN Ia
progenitors \citep[180 Myr according to][see \S~\ref{sec:Relevance}]{aubourg07:massive_stars_SNIa}, it is possible to
use the A+B model introduced in \S~\ref{sec:SNtoSNR:Ia} and the local SFHs to predict what fraction of Type Ia SNe will
come from prompt and delayed progenitors in each subcell ($f_{IaSN}$). For the prompt component, we have calculated an
average star formation rate by integrating the SFHs between 64 Myr \citep[the minimum time necessary to produce a CO WD
at the metallicity of the LMC,][]{dominguez99:Intermediate_Mass_Models} and 180 Myr, and we have multiplied it by the
value of $B$ from \citet{sullivan06:SNIa_rate_host_SFR}, $3.9\pm0.7\times10^{-4} \, \mathrm{SNe\, yr^{-1}\, (M_{\odot}\,
  yr^{-1})^{-1}}$. For the delayed component, we have simply multiplied the number of stars in each subcell by the value
of $A$ from \citet{sullivan06:SNIa_rate_host_SFR}, $5.3\pm1.1 \times 10^{-14} \, \mathrm{SNe\, yr^{-1}\,
  M_{\odot}^{-1}}$. The values of $f_{IaSN}$ for each Type Ia SNR are listed in Table \ref{tab-3}, and can be compared
with the values obtained for the whole LMC using the integrated SFH from Figure \ref{fig-2}: $f_{IaSN,Prompt}=59\%$ and
$f_{IaSN,Delayed}=41\%$. Even more so than in the case of CC SNe, we stress that these numbers should be considered with
caution, because they make strong assumptions about the properties of Type Ia SN progenitors. In particular, increasing
the upper limit to the age of the prompt component from 180 to 300 Myr results in an increase of the fraction of
\textit{delayed} Type Ia progenitors between 8 and 15 percentual points, depending on the SNR.

At a first glance, it is somewhat surprising that the SN Ia rates from prompt and delayed progenitors are always
comparable both in the whole LMC and at the locations of the individual Type Ia SNRs. This can be easily understood by
examining Figure 6 in \citet{sullivan06:SNIa_rate_host_SFR}: for low values of the SFR, the \textit{specific} rate of
Type Ia SNe from the prompt and delayed components is very similar, and low-intensity SF has been widespread across the
LMC during the last 5 Gyr (see \S~\ref{sec:Map} and HZ09).

\paragraph{SNR DEM L71} The exceptionally low rate of SF between 64 and 180 Myr in the subcell containing this SNR
makes it the most likely object ($72\%$ probability) to be associated with a delayed Type Ia SN
progenitor. Unfortunately, DEM L71 is also the oldest Type Ia SNR, and detailed information about its parent SN is hard
to extract from the observations. In particular, the ejecta emission is not remarkable in any way, and seems consistent
with normal Type Ia SN explosion models (BH09). The forward shock is running into material with a factor $\sim3$ density
range \citep{ghavamian03:DEML71}, but this can be easily explained by inhomogeneities in the ISM. The SNR dynamics
do not indicate any substantial modification of the AM by the progenitor \citep{badenes07:outflows}.

\paragraph{SNR N103B} Of all the SFHs associated with Type Ia SNRs that we discuss here, that of SNR N103B is without
doubt the most remarkable. This SNR is in a region of the LMC bar that has seen vigorous SF activity in the recent past,
with a prominent extended peak between 100 and 50 Myr \citep[probably associated to the nearby cluster NGC
1850,][]{gilmozzi94:NGC_1850} and a more recent one 12 Myr ago. The intensity of this last peak is even larger than
those associated with SN 1987A and SNR 158A in the 30 Dor region. It is not surprising that
\citet{chu88:LMC_SNRs_environments} mistyped SNR N103B as a CC SNR before a good quality X-ray spectrum was available,
based on this evident association with recent SF. This does not necessarily imply that the progenitor of SNR 103B was a
young star, although the predicted fraction of prompt Ia progenitors ($73\%$) is higher for this SNR than for any of the
others. However, if the properties of the SNR itself are taken into account, the association with recent SF becomes more
intriguing. \citet{lewis03:N103B} noted a number of similarities between the strong lateral asymmetry of SNR N103B and
that of Kepler's SNR. In the case of SNR N103B, it is unclear whether this asymmetry is due to an interaction with an
ISM structure (e.g., the nearby HII region DEM 84) or to some kind of CSM modified by the SN progenitor. In the Kepler
SNR, however, it has been shown that the asymmetry is indeed due to interaction with a CSM modified by the Type Ia SN
progenitor \citep[see][and references therein]{reynolds07:kepler}, suggesting that either the progenitor or its
companion must have been relatively massive. From the shape of the SFH, a relatively young (less than 150 Myr) and
metal-rich ($Z=0.008$) progenitor for SNR N103B seems a likely possibility.

\paragraph{SNR \ofo} This SNR is in a region that HZ09 called the Northwest Void due to its conspicuous lack of recent
star formation. In fact, HZ09 argue that the old and metal-poor stars in this part of the LMC are representative of the
primordial stellar population of the LMC. Moreover, the SNR is in a very homogeneous region (see Figure \ref{fig-3}),
with neighboring subcells having very similar stellar populations. The closest subcells with noticeable SF activity
(above $10^{-3}\,\mathrm{M_{\odot} yr^{-1}}$) at $t<100$ Myr are roughly 1 kpc away from the SNR. This would not be
noteworthy for a generic Type Ia SNR, but thanks to the work of \citet{rest08:0509} and \citet{badenes08:0509} we know
that SNR \ofo\ was originated by an exceptionally bright Type Ia SN which synthesized $\sim1\,\mathrm{M_{\odot}}$ of
$^{56}$Ni. This seems to be at odds with the conventional wisdom regarding prompt and delayed Type Ia
SN progenitors, which holds that exceptionally bright Type Ia SNe are usually associated with younger stellar
populations \citep{gallagher05:chemistry_SFR_SNIa_hosts}, but in fact the local SFH predicts equal contributions from
prompt and delayed progenitors even in this remarkably quiet part of the LMC. If this SNR did have a relatively young
and massive progenitor, however, it appears to have left the AM around it relatively undisturbed
\citep{badenes07:outflows,badenes08:0509}.

\paragraph{SNR \ofn} This SNR falls in line with \ofo\ and DEM L71 in having a low SFH at all times, although the local
stellar population appears to be more metal rich than in the other Type Ia SNRs (see Figure \ref{fig-7}). Other than
that, SNR \ofn\ is unremarkable both in its overall structure and dynamics and in the properties of its ejecta
emission. From the local SFH, delayed Type Ia progenitors are slightly favored over prompt ones, but not significantly.

\section{DISCUSSION}
\label{sec:Disc}

\subsection{Core Collapse Supernovae}

The combination of SNR studies and local SFHs that we have introduced in this paper is a promising method to further our
understanding of CC SN progenitors, but it needs to be refined before it can have a significant impact on this field of
research. Two major issues need to be addressed. First, the techniques used to determine the subtype of CC SNe from
their SNRs are still too crude to provide consistent results, even for well observed objects like the CC SNRs in our
sample. And second, the ability of tools like StarFISH to recover the SFH from mixed stellar populations at the ages
that are relevant for the evolution of CC SN progenitors (below 40 Myr) using photometric data needs to be firmly
established. It would be interesting to investigate the possibility of an increased temporal resolution at very early
times (less than 10 Myr) in order to distinguish between Type IIL/b and Type Ib/c SN progenitors, but this might require
support from spectroscopic surveys like the VLT-FLAMES \citep{evans06:VLT-FLAMES}. We hope to address these issues in
future publications.

Despite the limitations of the method, we have been able to obtian some interesting results. In general, we can say that
the properties of the SNRs and the local SFHs are compatible with each other, allowing for the large uncerainties
discussed above. Among our target objects, the SFH around SNR N49 seems to indicate that stars more massive than
$21.5\,\mathrm{M_{\odot}}$ are scarce in this part of the LMC. This opens two possibilities: either very massive
($>30\,\mathrm{M_{\odot}}$) stars are not necessary to produce magnetars, or the source SGR 0526$-$66 is not in fact
associated with SNR N49, as suggested by \citep{klose04:NIR_survey_N49_SGR0526-66}.  The SFH around SNR N63A seems
compatible with a very massive SN progenitor, maybe a Type Ib/c SN, but other possibilities cannot be discarded. For SNR
158A, the temporal resolution of the HZ09 maps is too coarse to resolve the issue of the progenitor. In any case, our
findings stress the importance of revisiting and reanalyzing the X-ray emission from these young CC SNRs in detail to
learn more about the SN explosions that originated them.

\subsection{Type Ia Supernovae}
\label{sec:Conc-Ia}

We have found that the combination of SNR studies and local SFHs is a powerful tool to explore the properties of Type Ia
SN progenitors. The X-ray emission from Type Ia SNRs is well understood in terms of explosion physics, and the ability
of SNR studies to recover the Type Ia SN subtype (i.e., bright vs. dim) has been demonstrated by
\citet{badenes08:0509}. Moreover, the combination of StarFISH and a data set like the MCPS is ideally suited to
characterize the stellar population in different parts of the LMC.

Among the Type Ia SNRs that we have studied, one (SNR N103B) is found in a region that has experienced vigorous SF in
the recent past, but the other three are associated to old and metal-poor stellar populations. On a first inspection, it
would be tempting to establish connections between these two kinds of environments and the prompt and delayed components
to Type Ia progenitors proposed by \citet{scannapieco05:A+B_models}, but we have found that the situation is more
complex than that. Even in regions with a remarkably small amount of recent SF, it is very hard to isolate objects that
arise unambiguously from delayed Type Ia progenitors. This is due to the high efficiency of the still unknown mechanisms
that turn CO WDs from relatively young, intermediate-mass stars into Type Ia SN progenitors of the prompt component
\citep{mannucci06:two_progenitor_populations_SNIa,pritchet08:SNIa_progenitors,maoz08:fraction_intermediate_stars_Ia_progenitors}. In
order to identify with some confidence a Type Ia SNR as a product of the prompt or delayed component, it would have to
be either in a region with very vigorous SF or in a pristine region with no measurable SF activity in the appropriate
range of ages. Since even elliptical galaxies appear to have some residual SF during their entire lifetimes
\citep{kaviraj08:recent_SF_ellipticals}, isolating the delayed Type Ia progenitors in the local universe to study their
properties in detail might be a very challenging task, unless a Type Ia SN is found in a nearby globular cluster
\citep{shara02:SNIa_in_star_clusters}.

Our results allow us to test specific theoretical predictions about Type Ia progenitors, like the claim by
\citet{kobayashi98:lowZ_inhibition_IaSNe} that the Type Ia SN rate should be very low for metallicities lower than a
tenth of the solar value. This is based on the so-called `accretion wind' scenario for single-degenerate Type Ia
progenitors, which requires a minimum opacity in the material transferred from the companion to the WD
\citep{hachisu96:progenitors}. The average \textit{stellar} metallicity is close to this value or even much lower around
three of the four Type Ia SNRs that we have examined, which seems hard to reconcile with the results of
\citeauthor{kobayashi98:lowZ_inhibition_IaSNe}, although we stress that all the regions that we examined do contain a
small number of stars with higher metallicities. Similar concerns about this prediction and its implications have been
raised by \citet{prieto08:SN_Progenitors_Metallicities}, who found several Type Ia SNe in low-metallicity dwarf
galaxies. The accretion wind scenario also makes strong predictions about the shape of the CSM around Type Ia
progenitors that are not substantiated by the dynamics of known Type Ia SNRs \citep{badenes07:outflows}. In this
context, an interesting possibility is opened by the recent work of \citet{badenes08:mntocr}, which allows one to make
\textit{direct} measurements of the metallicity of Type Ia SN progenitors using Mn and Cr lines in the X-ray spectra of
young SNRs. If this technique could be applied to the LMC SNRs, we would be able to contrast the results with the
properties of the stellar populations, and test theoretical ideas about the role of metallicity in different kinds of
Type Ia SNe \citep[e.g.][]{timmes03:variations_peak_luminosity_SNIa}.

Two of the SNRs we have examined have remarkable properties with important implications for Type Ia SN progenitors. The
unusual morphology of SNR N103B \citep[see][and references therein]{lewis03:N103B}, which is strongly suggestive of some
kind of CSM interaction, has become even more noteworthy in light of the vigorous recent SF revealed by the local
SFH. It appears that SNR N103B might be a member of an emerging class of Type Ia SNRs with CSM interaction that could be
associated with relatively young and massive progenitors that lose an appreciable amount of mass before exploding as
Type Ia SNe. This class would include the Kepler SNR \citep{reynolds07:kepler} in our Galaxy and other LMC SNRs
\citep{borkowski06:DEML238_DEML249}, but the local SFHs around these objects should be examined to confirm this
possibility. Evident signs of CSM interaction, however, cannot be found in other well studied Type Ia objects like Tycho
and SN 1006 in our own Galaxy or the other three LMC SNRs that we have analyzed here \citep{badenes07:outflows},
indicating that a majority of Type Ia progenitors do not modify their surroundings in a noticeable way. Since it is
unlikely that \textit{all} these other objects have had progenitors from the delayed component, we are left with two
possibilities: either the amount of mass loss during the pre-SN evolution of prompt Type Ia progenitors has a large
dynamic range or there is more than one way to produce Type Ia SNe with short delay times.

The properties of SNR \ofo\ are also remarkable, for entirely different reasons. \citet{rest08:0509} found $\Delta
m_{15}<0.9$ for this SN, which translates into to a $V$ magnitude at maximum light close to $-19.5$
\citep{phillips93:Ia-LLCcorrelation}. Yet, the SN exploded in a large region of the LMC with very little SF in the
recent past (see Figure \ref{fig-3}). The stars in the subcell that contains this SNR are on average
very old ($\bar{t_{*}}=7.9$ Gyr) and metal-poor ($\bar{Z_{*}}=0.0014$). \citet{gallagher08:SNIa_Early_Type_Galaxies} do
find some relatively bright Type Ia SNe associated with old stellar populations, but all their objects with peak $V$
magnitude above $-19$ and ages above 5 Gyr have large error bars on the age axis (see their Figure 5). Thus, SNR \ofo\
is probably the first \textit{bona fide} example of an exceptionally bright Type Ia SN associated with an old stellar
population. We note that our measurement of $\bar{t_{*}}$ should be very reliable, because it has not been derived from
a luminosity-weighted spectrum. It is important to stress that these bulk properties of the stellar population around
SNR \ofo\ do not preclude a relatively young progenitor for this object. During the age range that we have adopted for
prompt Type Ia progenitors, $2.1\times10^{4}\,\mathrm{M_{\odot}}$ of stars were formed in the subcell that contains SNR
\ofo. With a Salpeter IMF, roughly $10\%$ of this mass is in the 4 to 6 $\mathrm{M_{\odot}}$ range \citep[the ZAMS
masses that give CO WDs on timescales shorter than 180 Myr according to][]{dominguez99:Intermediate_Mass_Models}, which
results in a few hundred CO WDs from young stars. This number may seem small, but observational constraints on the
percentage of CO WDs that eventually explode as SN Ia are high \citep[2 to 40 \% according
to][]{maoz08:fraction_intermediate_stars_Ia_progenitors}, making a prompt progenitor for SNR \ofo\ a perfectly
reasonable possibility. The fact that an object like SNR \ofo\ appears in a sample of only four SNRs implies that bright
Type Ia SNe in old stellar populations may not be an exceptional occurrence, which should be taken into account when
examining the contribution of bright and dim Type Ia SNe in cosmological studies. 

Our results underline the dangers of trying to understand the behavior of Type Ia SN progenitors by studying only the
bulk properties of unresolved stellar populations in distant galaxies. If the LMC had been a distant Type Ia SN host,
two objects with such radically different SFHs as SNRs N103B and \ofo\ would have been assigned the same age and
metallicity. Even average quantities obtained from resolved stellar populations like $\bar{t_{*}}$ and $\bar{Z_{*}}$ can
be misleading if they are used by themselves to characterize the properties of SN progenitors - compare the values for
CC and Type Ia SNRs from Tables \ref{tab-2} and \ref{tab-3}.

\section{CONCLUSIONS}
\label{sec:Conc}

In this paper, we have presented the first systematic study of the stellar populations around CC and Type Ia SNRs in the
LMC. Our ultimate goal is to use all the available information on the X-ray emitting SNRs and the resolved stellar
populations of the Magellanic Clouds to improve our understanding of CC and Type Ia SN progenitors, their final
evolutionary stages, the SN explosions that mark their demise, and the aftermath of these explosions. In that broad
context, this work only represents a first exploration of the many possibilities that are opened by recent theoretical
and observational advances in both SNR research and stellar population studies. We plan to pursue this line of research
in the future, increasing the sample of objects and refining the techniques that we have presented here.

We have found that the local SFHs around the CC SNRs in our sample (N49, SN 1987A, N63A, and N158A) are always dominated
by significant episodes of SF in the recent past ($t<40$ Myr), as expected from previous observational and theoretical
work. The timing and intensity of these SF episodes can provide interesting constraints on the masses of CC SN
progenitors, but more work is needed to explore the full potential of this method.

The local SFHs have also allowed us to study the ages and metallicities of the stellar populations around our target
Type Ia SNRs (DEM L71, N103B, \ofo, and \ofn) in great detail. We have found that Type Ia SNe explode in a variety of
environments, ranging from old and metal-poor populations to sites with vigorous SF in the recent past. Using the
two-component model proposed by \citet{scannapieco05:A+B_models}, we have explored the relationship between specific
properties of Type Ia SNe and their parent stellar populations. We have seen that extremely bright Type Ia SNe can
explode very far away from any significant star formation activity (SNR \ofo), and that Type Ia SNe associated with
young stellar populations might sometimes experience significant mass-loss before they explode (SNR N103B). Recent
studies of extragalactic Type Ia SN rates and our own findings suggest that reality is probably too complex to be
explained with the popular two-component progenitor model. If this is so, high-quality SFHs for Type Ia SNRs obtained
from resolved stellar populations like the ones we present here should provide an excellent observational constraint on
new ideas about Type Ia progenitors.

\acknowledgements This work has benefited from many conversations with a large number of colleagues, including Steve
Bickerton, Jeremy Goodman, Jim Gunn, Jack Hughes, Raul Jim\'{e}nez, Vicky Kaspi, Dan Maoz, Maryam Modjaz, Evan
Scannapieco, Jerry Sellwood, and Kris Stanek. We are also grateful to the referee, Stephen Smartt, for many helpful
suggestions that improved the quality of our manuscript. Support for this work was provided by NASA through
\textit{Chandra} Postdoctoral Fellowship Award PF6-70046 issued by the \textit{Chandra} X-ray Observatory Center, which
is operated by the Smitsonian Astrophysical Observatory for and on behalf of NASA under contract NAS8-0360. DZ
acknowledges support from NASA LTSA grant NNG05GE82G. JLP is supported by NSF grant AST-0707982.


\bibliographystyle{apj} 

\clearpage

\clearpage

\begin{center}
  \begin{deluxetable}{lcccccc}
    \tablewidth{0pt}
    \tabletypesize{\scriptsize}
    \tablecaption{Young Supernova Remnants in the Large Magellanic Cloud \label{tab-1}}
    \tablecolumns{7}
    \tablehead{
      \colhead{} &
      \colhead{Common} &
      \multicolumn{2}{c}{Position (J2000)  \tablenotemark{b}} &
      \colhead{Size} &
      \colhead{Age} &
      \colhead{Age Estimation}\\
      \colhead{SNR \tablenotemark{a}} &
      \colhead{Name} &
      \colhead{RA} &
      \colhead{Dec} &
      \colhead{(arcmin)} &
      \colhead{(yr)} &
      \colhead{Method}
    }
    
    \startdata
    0505$-$67.9 & DEM L71 & 05h 05m 42s & -67d 52m 39s & 1.4x1.0 (H03) & $4360 \pm 290$  (G03) & SNR Dynamics \tablenotemark{c} \\
    0509$-$68.7 & N103B & 05h 08m 59s & -68d 43m 35s & 0.46 (B07) & $\sim$860 (R05) &  Light Echo\\
    \ofo\ &  & 05h 09m 31s & -67d 31m 17s & 0.48 (B07) & $400 \pm 50$ \tablenotemark{d} (B08,R05) & Light Echo \\
    \ofn\ &  & 05h 19m 35s & -69d 02m 09s & 0.52 (B07) & $600 \pm 200$ (R05) & Light Echo \\
    0525$-$66.1 & N49 & 05h 26m 00s & -66d 04m 57s & 1.4 (P03) & $6300 \pm 1000$  (V06) & SNR Dynamics \tablenotemark{c} \\ 
    SN 1987A & SN 1987A & 05h 35m 28s & -69d 16m 11s & 0.03 (N08) & 22 & SN \\
    0535$-$66.0 & N63A & 05h 35m 44s & -66d 02m 14s & 1.1 (W03) & $3500 \pm 1500$ (H98) & SNR Dynamics \tablenotemark{c} \\
    0540$-$69.3 & N158A & 05h 40m 11s & -69d 19m 55s & 1.3x0.7 (H01) & $\sim$800 (C05) & Pulsar 
    \enddata

    \tablenotetext{a}{By convention, LMC SNRs are designated in this abbreviated form using the RA and Dec of their
      center in J1950 coordinates. For more details on the SNR names, see \citet{williams99:LMC_SNR_Atlas}.}    

    \tablenotetext{b}{From \citet{williams99:LMC_SNR_Atlas}.}

    \tablenotetext{c}{Ages estimated from SNR dynamics are subject to substantial uncertainties. See text for details.}

    \tablenotetext{d}{The age from the light echo dynamics is $400 \pm 120$ yr \citep{rest05:LMC_light_echoes}, but the
      spectral and dynamical properties of the SNR, together with some historical considerations, constrain the value
      much more, see discussion in \S~5.3 of \citet{badenes08:0509}.}

    \tablerefs{B07: \citet{badenes07:outflows}; B08: \citet{badenes08:0509}; C05: \citet{chevalier05:young_cc_SNRs};
      G03: \citet{ghavamian03:DEML71}; H01: \citet{hwang01:N158A}; H03: \citet{hughes03:DEML71}; 
      H98: \citet{hughes98:LMC_SNRs_ASCA}; N08: \citet{ng08:SN1987A}; P03: \citet{park03:N49}; 
      R05:\citet{rest05:LMC_light_echoes}; R08: \citet{rest08:0509}; V06: \citet{vink06:SNRs_Magnetars};
      W03: \citet{warren03:N63A}}
    
  \end{deluxetable}
\end{center} 

\begin{center}
  \begin{deluxetable}{lccccccc}
    \tablewidth{0pt}
    \tabletypesize{\scriptsize}
    \tablecaption{Core Collapse Supernova Remnants \label{tab-2}}
    \tablecolumns{8}
    \tablehead{
      \colhead{} &
      \colhead{CC Typing} &
      \colhead{Subtype} &
      \colhead{} &
      \colhead{$\bar{t_{*}}$ \tablenotemark{b}} &
      \multicolumn{3}{c}{$f_{CCSN}$ (in \%)}\\
      \colhead{SNR} &
      \colhead{Criteria \tablenotemark{a}} &
      \colhead{classification} &
      \colhead{$\bar{Z_{*}}$ \tablenotemark{b}} &
      \colhead{(Gyr)} &
      \colhead{$8$ to $12.5\,\mathrm{M_{\odot}}$} &
      \colhead{$12.5$ to $21.5\,\mathrm{M_{\odot}}$} &
      \colhead{$>21.5\,\mathrm{M_{\odot}}$}
    }
    

    \startdata
    N49 & OB (C88), CO? (G01) & Unknown (IIP?, B07) & 0.0018 & 6.0 & $0$ & $100$ & $0$ \\ 
    SN 1987A & SN & 1987-like/IIpec & 0.0043 & 8.7 & $20$ & $56$ & $24$ \\
    N63A & OB (C88) & Unknown (Ib/c?, H98) & 0.0013 & 6.4 & $0$ & $30$ & $70$ \\ 
    N158A & CO (K89) & IIP (W08) or Ib/c (C05) & 0.0036 & 7.8 & $30$ & $36$ & $34$
    \enddata
  
    \tablenotetext{a}{OB: in or very close to an OB association. CO: compact object. SN: SN spectroscopy.}

    \tablenotetext{b}{These values are provided for comparison with the Type Ia SNe listed on Table \ref{tab-3}, and
      do not reflect the properties of the progenitor stars of the CC SNRs.}

    \tablerefs{B07: \citep{bilikova07:N49};C88: \citet{chu88:LMC_SNRs_environments}; C05: \citet{chevalier05:young_cc_SNRs}; K89:
      \citet{kirshner89:0540}; G01: \citet{gaensler01:AXP_SGR_SNR}; H98: \citet{hughes98:LMC_SNRs_ASCA}; W03:
      \citet{warren03:N63A}; W08: \citet{williams08:0540}}
    
  \end{deluxetable}
\end{center}

\begin{center}
  \begin{deluxetable}{lcccccc}
    \tablewidth{16cm}
    \tabletypesize{\scriptsize}
    \tablecaption{Type Ia Supernova Remnants \label{tab-3}}
    \tablecolumns{7}
    \tablehead{
      \colhead{} &
      \colhead{Ia typing} &
      \colhead{Subtype} &
      \colhead{} &
      \colhead{$\bar{t_{*}}$} &
      \multicolumn{2}{c}{$f_{IaSN}$ (in $\%$)}\\
      \colhead{SNR} &
      \colhead{Criteria} &
      \colhead{Classification \tablenotemark{a}} &
      \colhead{$\bar{Z_{*}}$} &
      \colhead{(Gyr)} &
      \colhead{Prompt} &
      \colhead{Delayed}
    }
    
    \startdata
    DEM L71 & X-ray (H95) & Normal? (BH09) & 0.0022 & 8.3 & 28 & 72 \\
    N103B & X-ray (H95) & Bright? (BH09) & 0.0023 & 8.1 & 73 & 27 \\
    \ofo\ & X-ray (H95) & Very bright (B08,R08) & 0.0014 & 7.9 & 49 & 51\\
    \ofn\ & X-ray (H95) & Bright (BH09) & 0.0032 & 7.7 & 41 & 59 
    \enddata
    
    \tablenotetext{a}{The subtype corresponds to the estimated brightness of the SN inferred from the mass of $^{56}$Ni
      in the best-fit model for the X-ray spectrum of the SNR: Very bright ($\sim1\,\mathrm{M_{\odot}}$ of $^{56}$Ni),
      bright ($\sim0.8\,\mathrm{M_{\odot}}$ of $^{56}$Ni) normal ($\sim0.6-0.4\,\mathrm{M_{\odot}}$ of $^{56}$Ni), and
      dim ($\sim0.3\,\mathrm{M_{\odot}}$ of $^{56}$Ni).}
  
    \tablerefs{B08: \citet{badenes08:0509};
      BH09: \citet{badenes09:IaSNRs_LMC};
      H95: \citet{hughes95:typing_SN_from_SNR};  
      R08: \citet{rest08:0509}}
    
  \end{deluxetable}
\end{center}

\end{document}